\documentclass[%
 aip,%
 jcp,%
 amsmath,amssymb,%
preprint,%
citeautoscript,%
]{revtex4-2}

\usepackage{graphicx}
\usepackage{dcolumn}
\usepackage{bm}
\usepackage[version=4]{mhchem}
\usepackage{xcolor}
\usepackage{siunitx}
\DeclareSIUnit\angstrom{\text {Å}}
\usepackage{braket}
\usepackage{xr}
\usepackage{balance}

\begin{document}

%
%

\title{Exciton--photocarrier interference in mixed lead-halide-perovskite nanocrystals}

\author{Esteban~Rojas-Gatjens}
\affiliation{School of Chemistry and Biochemistry, Georgia Institute of Technology, 901 Atlantic Drive, Atlanta, GA~30332, United~States}
\affiliation{School of Physics, Georgia Institute of Technology, 837 State St NW, Atlanta, GA~30332, United~States}

\author{Quinten~A.~Akkerman}
\altaffiliation{Current Affiliation: Chair for Photonics and Optoelectronics, Nano-Institute Munich and Department of Physics, Ludwig-Maximilians-Universit\"at (LMU), K\"oniginstra{\ss}e 10, 80539 Munich, Germany}
\affiliation{Nanochemistry Department, Istituto Italiano di Tecnologia, Via Morego 30, 16163 Genova, Italy}

\author{Liberato~Manna}
\affiliation{Nanochemistry Department, Istituto Italiano di Tecnologia, Via Morego 30, 16163 Genova, Italy}

\author{Ajay~Ram~Srimath~Kandada}
\email{srimatar@wfu.edu}
\affiliation{~Department of Physics and Center for Functional Materials, 2090 Eure Drive, Wake Forest University, Winston-Salem, NC~27109, United~States}

\author{Carlos~Silva-Acu\~na}
\email{carlos.silva@umontreal.ca}
\affiliation{School of Chemistry and Biochemistry, Georgia Institute of Technology, 901 Atlantic Drive, Atlanta, GA~30332, United~States}
\affiliation{School of Physics, Georgia Institute of Technology, 837 State St NW, Atlanta, GA~30332, United~States}
\affiliation{Institut Courtois \& D\'epartement de physique, Universit\'e de Montr\'eal, 1375 Avenue Th\'er\`ese-Lavoie-Roux, Montr\'eal, Qu\'ebec H2V~0B3, Canada}

\date{\today}

\begin{abstract}
The use of semiconductor nanocrystals in scalable quantum technologies requires characterization of the exciton coherence dynamics in an \emph{ensemble} of electronically isolated crystals in which system-bath interactions are nevertheless strong. 
In this communication, we identify signatures of Fano-like interference between excitons and photocarriers in the coherent two-dimensional photoluminescence excitation spectral lineshapes of mixed lead-halide perovskite nanocrystals in dilute solution. 
Specifically, by tuning the femtosecond-pulse spectrum, we show such interference in an intermediate coupling regime, which is evident in the coherent lineshape when simultaneously exciting the exciton and the free-carrier band at higher energy. 
We conclude that this interference is an intrinsic effect that will be consequential in the quantum dynamics of the system and will thus dictate decoherence dynamics, with consequences in their application in quantum technologies.
\end{abstract}

\maketitle

%
%

\section{Introduction \label{sec:intro}}

In condensed matter, the role of the noisy bath in dissipating the coherence induced by system-light interactions is of fundamental significance in dictating the optical properties of materials. The \textit{bath} is often understood to imply electron-phonon coupling~\cite{Rudin1990}, dielectric fluctuations in a noisy thermal environment, or inter-particle Coulomb scattering~\cite{Li2006, gregoire2017excitonic, Srimath2020Stochastic}, for example; here we consider the effect of interference of multiple close-lying optical states, which can be an intrinsic source of decoherence arising from the specific electronic structure of the material.
There are several examples of inter-excitonic coherence in II-VI colloidal quantum dots~\cite{Turner2012_Superposition, Caram2014, Collini2019_JPCC}. 
For example, Cassette~\textit{et al.}\ measured the coherent superposition of heavy-hole and light-hole excitons in \ce{CdSe} nanoplates~\cite{Cassette2015}.
Quantifying these types of inter-state coherent coupling and the processes leading to the loss of coherence is particularly important in the development of materials for quantum technologies, where the goal is to harness and manipulate electronic coherence in photoexcitations~\cite{Collini2019_JPCC, Kagan2021_ChemicalReviews}. 
The implementation of semiconductor nanocrystals (NCs), specifically in a quantum confinement regime (quantum dots, in which the crystal dimension is smaller than the exciton Bohr radius), in quantum-technological applications, such as single-photon sources and photonic quantum simulation, is of high contemporary interest~\cite{Utzat2019, Tao2022, Chen2022}. 
Significantly, the optical coherence time of emitted photons in single metal-halide perovskite NCs was determined to be in the order of picoseconds, comparable to the radiative lifetime~\cite{Utzat2019},
positioning perovskite NCs as promising quantum emitters.  
In these contexts, it is important to understand the complex quantum dynamics to both minimize decoherence and exploit scattering mechanisms as a handle of nonlinear interactions in matter~\cite{Chang2014}. 

A tool of choice to measure decoherence dynamics is coherent nonlinear spectroscopy, but such reports 
in perovskite NCs are scarce~\cite{Seiler2019, Yu2021, Liu2021}, in contrast to the case of II-VI semiconductor quantum dots~\cite{Turner2012_Superposition, protesescu2015nanocrystals, Cassette2015}. 
In the latter systems, the photophysics involves discrete and well-resolved excited-state resonances~\cite{Bawendi1990}. 
Based on incoherent nonlinear spectroscopy, Butkus~\textit{et al.} demonstrated that in pure metal-halide perovskite cubic NCs, strong quantum confinement signatures manifest in transient absorption lineshapes only when the edge length is smaller than $\sim 4$\,nm~\cite{Butkus2017}.  
As a consequence, they concluded the spectral signatures of metal-halide perovskite NCs with size $\gtrsim 4$\,nm resemble those of the bulk semiconductor.
More recently, the synthesis of monodisperse spheroidal perovskite NCs results in well-resolved higher energy excitonic signatures for crystal sizes of 10\,nm~\cite{Quiten2022, barfuser2022}.

In this communication, we measure the coherent nonlinear response of mixed-lead-halide perovskite NCs utilizing coherent two-dimensional photoluminescence excitation spectroscopy (2D-PLE).
The NCs used in this study do not exhibit strong quantum confinement effects. 
Their photophysical characteristics are akin to their bulk counterparts. However, the major difference lies in the reduced defect density. We have previously observed that the 2D-PLE technique fails to unambiguously measure the coherent response in bulk perovskite films, due to the incoherent mixing of photo-generated population operating in the defect-limited regime, see Ref.s~\citenum{gregoire2017incoherent, bargigia2022identifying}. 
Such incoherent contributions are effectively negated in NCs due to a substantial reduction in defect density, as evidenced by the intensity independent, and high photoluminescence quantum yield as well as by intensity independent PL dynamics. 
This ensures that we can reliably retrieve the true coherent nonlinear response of the material through 2D-PLE.  

We present evidence of coherent coupling between an exciton state and the free-carrier continuum via lineshape analysis of the complex coherent spectrum as a function of excitation energy.  
The nonlinear optical lineshape is well described by a model for Fano-like interference introduced by Finkelstein \textit{et al}~\cite{finkelstein2016coherent}. 
We propose that this is a general effect in the optical response of semiconductor NCs in scenarios in which the exciton resonance and the edge of the free-carrier band are nearly degenerate.

\section{Results and Discussion\label{sec:results}}

\begin{figure}
    \centering
    \includegraphics[width=8.5cm]{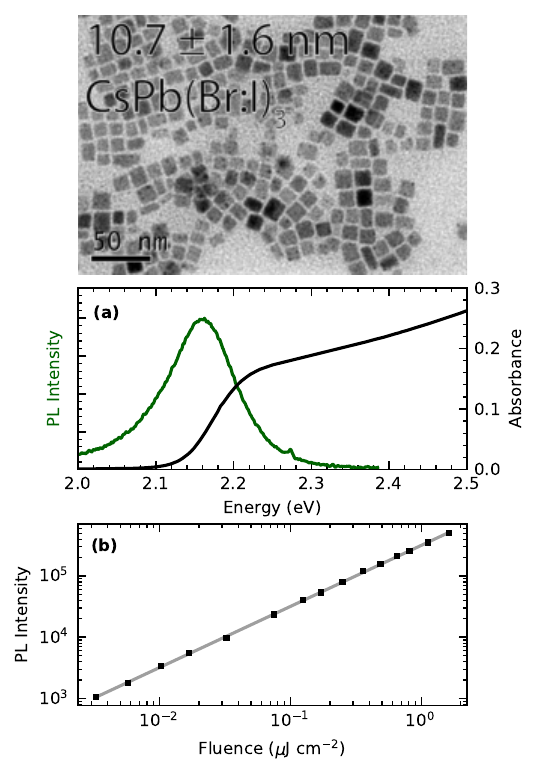}
    \caption{TEM image of an ensemble of CsPb(Br:I)$_3$ NCs, showing a cubic edge dimension of $10.7 \pm 1.6$\,nm. (a) Absorption and photoluminescence (PL) spectra of NC solution in toluene. 
    (b) Femtosecond-pulse fluence $\varphi$ dependence of the PL intensity, $I  = m\varphi^{\alpha}$, with $m = (3.2 \pm 0.1) \times 10^5\,\mu$J$^{-\alpha}$\,cm$^{2\alpha}$ and $\alpha = 0.998\pm 0.014$.} 
    \label{fig:AbsorptionPL}
\end{figure}

We studied CsPb(Br:I)$_3$ NCs with an average edge length of $10.7 \pm 1.6$\,nm, determined by transmission electronic microscopy (TEM, Fig.~\ref{fig:AbsorptionPL}). 
This size is slightly above that for quantum confinement~\cite{Protesescu2015}, placing this sample in a weak to intermediate confinement regime~\cite{Butkus2017, Zhao2020}. 
The reports of binding energy in bulk lead-halide perovskite ranges between 7 and 40\,meV~\cite{Hansen2024}, due to the weak confinement we expect the nanocrystals studied here to have similar binding energy.
Accordingly, the absorption spectrum shows a spectral lineshape similar to that of the bulk semiconductor (Fig.~\ref{fig:AbsorptionPL}(a))~\cite{Saba2014}, where the exciton absorption feature is not discernible, due to the low binding energy together with the homogenous and inhomogenous broadening.
The photoluminescence intensity scales linearly with excitation density, over two orders of magnitude, as shown in Fig.~\ref{fig:AbsorptionPL}(b).
The PL decay dynamics are also intensity-independent, albeit having a multi-exponential behavior possibly due to intrinsic sample inhomogeneity, (Fig.~S1). 
These observations indicate the recombination process is predominantly radiative with negligible contribution from defects.
Defect-assisted recombination results in a super-linear increase in the PL intensity with increasing excitation density, owing to the availability of a greater fraction of photo-generated carriers for radiative recombination at higher densities, see Refs.~\citenum{Srimath2016ECPL, Rojas2023ECPL}.
More pertinently to this work, given the PL response is linear, any nonlinearity measured through 2D-PLE over the same fluence range, (Fig.~\ref{fig:AbsorptionPL}(b)), is entirely due to the coherent response of the material. 
This is also confirmed by the response analysis shown in the SI and developed in previous work~\cite{bargigia2022identifying}.

\begin{figure}
    \centering
    \includegraphics[width=15cm]{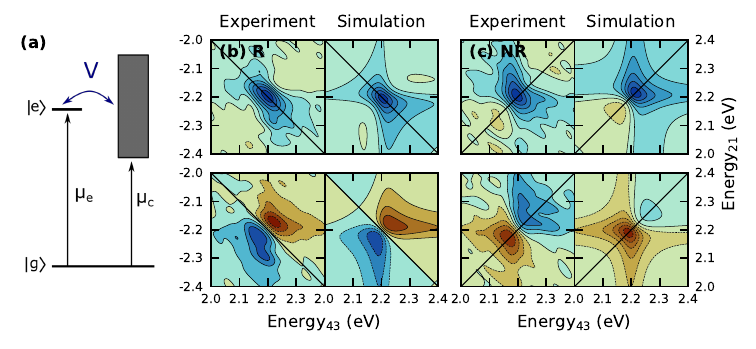}
    \caption{ (a) Diagrammatic representation of 
    the Fano inference scenario, where a discrete state couples to a continuum of states. 
    Experimental (left column) and simulated (right column, following equation~\eqref{Response} with $\gamma_c = \gamma_e = 0.02$\,eV, $\epsilon_e = 2.2$\,eV and $q=2$}) 2D-PLE spectra with (b) rephasing and (c) non-rephasing phase pathways. For each response, the top row is the real part and the bottom row is the imaginary part of the complex spectrum. The experimental femtosecond pulse spectrum is displayed in the top panel of Fig.~\ref{fig:spectraDependence}(b). 
    \label{fig:ExpSim}
\end{figure}

We performed 2D-PLE measurements in a dilute solution of CsPb(Br:I)$_3$ NCs at room temperature and analyzed the measured lineshapes in the context of a Fano interference model (Fig.~\ref{fig:ExpSim}(a)). 
For a single excitonic transition, we expect the real component of the rephasing spectrum to be symmetric along the anti-diagonal and the diagonal.  
In the real part of the rephasing 2D-PLE spectrum (Fig.~\ref{fig:ExpSim}(b), top left), we observe an asymmetry along the diagonal, and at higher energies. 
The asymmetry is also observed, in distinct form, in the imaginary components and the non-rephasing spectrum (Fig.~\ref{fig:ExpSim}(c), bottom right).
Similar asymmetric two-dimensional spectral lineshapes have been theoretically predicted~\cite{finkelstein2016coherent} and experimentally observed in a variety of material systems~\cite{Gandman2017, Roeding2018, Rojas-Gatjens_2024}. 
Fundamentally, it is attributed to the Fano interference between a discrete excited state and a continuum of states that energetically overlap. 
While this phenomenon is more common in the context of photonics and plasmonics~\cite{Miro2010, Luk2010}, in the case of semiconductors the presence of overlapping electronic excited states is suggested to lead to a similar scenario.
More precisely, here we interpret the spectral lineshape as the result of Fano-like interference of the \textit{discrete} excitonic state and the free-carrier continuum band. 

In this context, Finkelstein~\textit{et al.}\ derived analytical expressions for the two-dimensional coherent Fano-profiles~\cite{finkelstein2015fano, finkelstein2016coherent, Finkelstein2018FanoESA}.  
The Fano Hamiltonian, equation~\eqref{eq:FanoH}, describes the scenario for a discrete transition to state $\ket{e}$ with energy ($\epsilon_e$) coherently coupled to a continuum of states \{$\ket{\ell}$\} with energies \{$\epsilon_{\ell}$\} with a coupling constant $V$:

\begin{align}
    \begin{split}
        H 
          = \epsilon_g \ket{g}\!\bra{g} &+ \epsilon_e \ket{e}\!\bra{e} +\int_0^{\infty} \mathrm{d}\ell\,\epsilon_{\ell} \ket{\ell}\!\bra{\ell}  \\
    &+ \int_0^{\infty} \mathrm{d}\ell\,\left[V\,\ket{e}\!\bra{\ell}+V^*\,\ket{\ell}\!\bra{e} \right],
    \end{split}
    \label{eq:FanoH}
\end{align}
where $\ket{g}$ represents the ground-state wavevector with energy $\epsilon_g$. 

The expression for the rephasing response for the ground-state bleach contribution ($R_3$), obtained for the Hamiltonian described in equation~\eqref{eq:FanoH}, is shown in equation~\eqref{Response},~\footnote{We have ignored the population time dependence in equation~\eqref{Response} as we do not explore that variable in this work, however, the original derivations in ref.~\citenum{finkelstein2016coherent} include such dependence.} which assumes impulsive light-matter interactions, and also assumes that the continuum parameters are independent of energy:  

\begin{equation}
    R_3(\epsilon_{21}, \epsilon_{43}) \propto \left[\frac{\Gamma (q-i)^2}{\epsilon_{43}+i}-i\right]\left[\frac{\Gamma (q+i)^2}{\epsilon_{21}-i}+i\right],
    \label{Response}
\end{equation}
where $\Gamma = {\gamma_e}/({\gamma_e+\gamma_c})$ with $\gamma_e$ and $\gamma_c$ corresponding to the dephasing of the discrete and continuum states, and $\epsilon_{ij}$ are the dimensionless energy variables represented in Fig.~\ref{fig:ExpSim}. 
The parameter $q \equiv \mu_e / n\pi V\mu_c$ is the asymmetry factor grouping together the parameters associated with the discrete and continuum-state coupling, $\mu_e$ and $\mu_c$ are the transition dipole moments for the discrete and continuum states, respectively, and $n$ is the density of states in the continuum. 
Note that for very large $q$ ($\mu_e \gg n\pi V\mu_c$) the Lorentzian lineshape is recovered. 
The corresponding non-rephasing response ($R_4$) reads as:

\begin{equation}
R_4(\epsilon_{21}, \epsilon_{43}) \propto \left[\frac{\Gamma (q-i)^2}{\epsilon_{43}+i}-i\right]\left[\frac{\Gamma (q-i)^2}{\epsilon_{21}+i}-i\right].
  \label{NR_response}
\end{equation}

In the non-rephasing spectra, both axes are given by $\epsilon_{ij}(\omega) = (\omega - \epsilon_e/\hbar )/(\gamma_e+\gamma_c)$ and in the rephasing spectra the y-axis, which in the convention used in this work is the ``pump'' energy, is given by $\epsilon_{21}(\omega) = (\omega + \epsilon_e/\hbar )/(\gamma_e+\gamma_c)$.
The expressions for the stimulated emission pathways can be found in the original text~\cite{finkelstein2016coherent} and the Supporting Information.

\begin{figure}
    \centering
    \includegraphics[width=8.5cm]{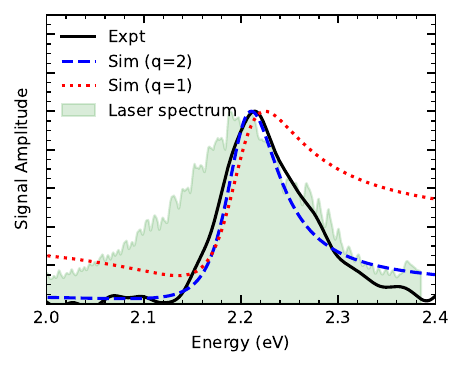}
    \caption{Diagonal cut of the real (absorptive) part of the 2D-PLE total correlation spectrum, obtained by summing the real part of the rephasing and non-rephasing components, $\Re \{I_{\mathrm{tot}}(\hbar \omega_{21}, \hbar \omega_{43})\} = \Re \{I_{\mathrm{R}}(\hbar \omega_{21}, \hbar \omega_{43})\} + \Re \{I_{\mathrm{NR}}(\hbar \omega_{21}, \hbar \omega_{43})\}$. We compare the experimental data (black solid line) and the simulation with two distinct coupling conditions, $q=1$ (dotted red line) and $q=2$ (dashed blue line), as defined in equation~\eqref{Response}.}
    \label{fig:lineshape}
\end{figure}

\begin{figure}
    \centering
    \includegraphics[width=15cm]{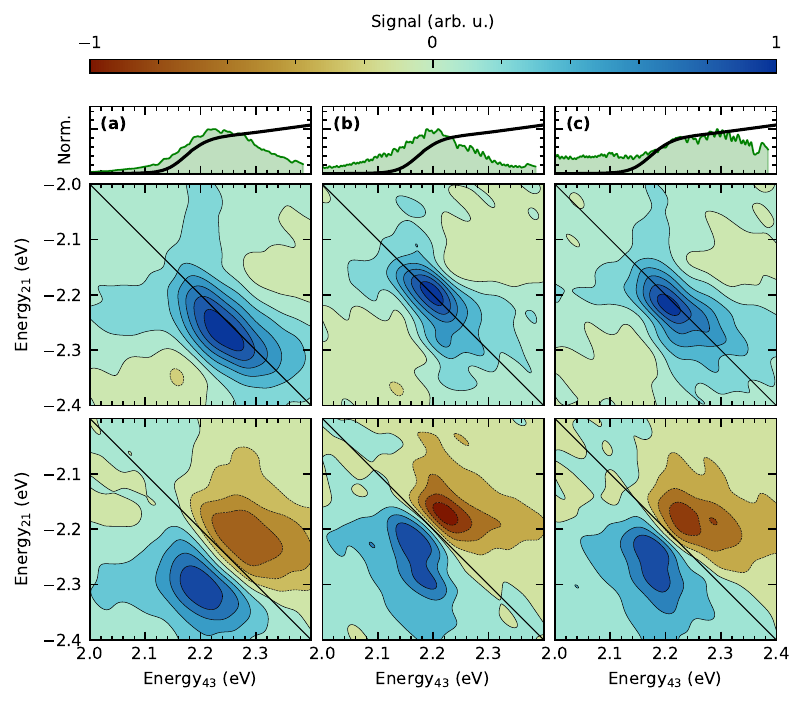}
    \caption{Absorption spectrum superimposed with the excitation-laser spectrum (top row), the real component (middle row), and the imaginary component (bottom row) of the rephasing spectrum for a laser bandwidth that is tuned to be (a) resonant with the semiconductor band-edge, (b) blue-shifted by $\sim 40$\,meV to cover a broader spectral range across the band-edge, and (c) broadened to increase the spectral range coverage across the band-edge. 
    }
    \label{fig:spectraDependence}
\end{figure}


The simulated nonlinear lineshapes agree well with the experimental lineshapes. Here, we considered 
the ground-state bleach and stimulated emission pathways,~\footnote{The excited-state pathways are not considered in this work as we do not observe any indication in the experimental data. To the interested reader, the derived expressions can be found in ref.~\citenum{Finkelstein2018FanoESA}.} and set $\gamma_c = \gamma_e = 0.02$\,eV, $\epsilon_e = 2.2$\,eV and $q=2$ (Fig.~\ref{fig:ExpSim}). 
Coherent nonlinear spectroscopy of metal halide perovskites has been performed and reported by several groups~\cite{Seiler2019, Yu2021, Brosseau2023}. Interestingly, these earlier works do not report any asymmetry in the spectral lineshapes.
We highlight that most of the analysis so far has been performed on the total correlation spectrum, which is the cumulative sum of the rephasing and non-rephasing components. 
In the present case, while we observe a distinct Fano-like lineshape in the rephasing and non-rephasing components, it is less evident in the absorptive components of the total correlation spectrum, consistent with the previous reports. 
We take a closer inspection of the diagonal cut of the total absorptive spectrum, shown in Fig.~\ref{fig:lineshape}. 
We show the simulated spectra obtained using equations~\eqref{Response}, \eqref{NR_response}, (S1), and (S2), and considering two coupling parameters set at $q=1$ and $q=2$, together with a diagonal cut of the experimental total absorptive component. 
The moderate exciton-continuum coupling defined by $q=2$ accurately resembles the experimental spectrum.
In fact, from linear spectroscopy, where Fano-profiles are usually identified, it would not have been possible to assign coherent exciton-carrier coupling in this system due to the subtlety of the effect under the coupling conditions reflected by this analysis. 

\begin{figure}
    \centering
    \includegraphics[width=8.5cm]{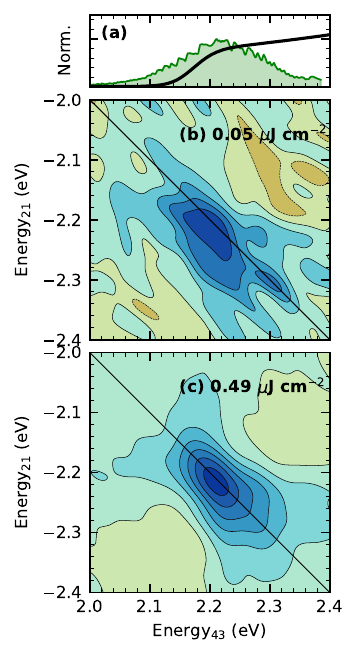}
    \caption{(a) Absorption spectrum superimposed with the excitation-laser spectrum (top row). We show the real component of the rephasing spectra measured with a fluence of (b) $0.05\,\mu J cm^{-2}$ and (c) $0.49\,\mu J cm^{-2}$.}
    \label{fig:PowerDep}
\end{figure}

To further substantiate the Fano interference of exciton and carrier continuum, we analyze the nonlinear spectra obtained with three distinct laser excitation spectra (Fig.~\ref{fig:spectraDependence}). 
We show the cases for (i) a laser spectrum that excites resonantly the free-carrier band only (Fig.~\ref{fig:spectraDependence}.a), (ii) a redshifted spectrum that excites the exciton and a part of the continuum close to the optical edge, and (iii) a broadband excitation pulse covering both the previous conditions. 
When pumping primarily the free-carrier band, we observe a spectrum expected of a continuum, which is inhomogeneously broadened, and importantly, without any signatures of Fano interference. 
Under condition (ii), we observe the signatures of the Fano-interference as the discrete state is also coherently excited, which, we suggest, interferes with the continuum of free carriers. 
To discard the case of two independent transitions (not interfering) that spectrally overlap, the third condition excites, with a broad spectrum, the band-edge and the high energy free-band states of the nanocrystals.
Note that when exciting two distinct transitions, the laser spectrum enhances the nonlinear responses from transitions resonant to it. 
We do not observe a significant change in the line shape supporting the interpretation that the asymmetric lineshape is caused by the interaction of the exciton at the band-edge and the free-carriers band that are coherently excited.
The pure dephasing width ($\gamma$), estimated from the full-width half max (FWHM) of the anti-diagonal cut of the absolute rephasing component ($2\gamma$) at 2.25\,eV, is $67 \pm 2$\,meV. 
Under conditions (ii) and (iii) $\gamma = 43 \pm 2$ and $46 \pm 2$\,meV at 2.2\,eV, respectively. The nonrephasing components of the spectra are shown in Fig.~S4.

As perovskite NCs have a shallow defect distribution within the band edge~\cite{Akkerman2018NatMat}, we performed intensity-dependent experiments with the idea that if the discrete state observed in the nonlinear lineshape corresponds to a shallow defect, we would see an intensity-dependent nonlinear lineshape. 
We do not observe a significant shift in the energies or spectra lineshape in the fluence range probed.
However, when comparing the spectra measured at the lowest fluence (0.05~$\mu$J$\,$cm$^{-2}$) with a spectrum measured at four times the power, we note that there are unresolved dynamics that lead to an asymmetry along the anti-diagonal in the lineshape.
We believe the spectral lineshape at low fluences can be associated with the interaction of the carriers with defect states.
Since the asymmetry of the cross peak is not observed anymore as we increase the excitation density we believe it can be related to the fast population transfer from high energy carriers to shallow defect states close to the band edge.
We measured the 2D-PLE spectra for a fluence range of 0.05-0.5~$\mu$J$\,$cm$^{-2}$ and we only observed the strong asymmetry in the cross peaks at the lowest excitation fluence, hinting that shallow defects saturate and no longer participate in the population dynamics.
We show the complete spectra in the Supplementary information~(Fig.~S5-9).
The thorough description of the population dynamics is not the main subject of this communication.

\section{Summary and outlook \label{sec:concl}}

In this communication, we have presented an analysis of the two-dimensional spectral response of colloidal nanocrystals of metal halide perovskites. 
Typically, the optical absorption spectra of these moderately confined nanocrystals are rationalized within the Elliott formalism with distinct excitonic resonances below the band-edge.
The nonlinear optical spectra discussed in this manuscript however manifest lineshapes that are indicative of coherent coupling between the exciton and free carrier continuum. Within the spirit of the Fano model of coupling of a discrete state and a continuum, and based on our observation, the optical excitations here must be perceived as mixed quantum states of excitons and free carriers.

Fano-interference necessitates an energetic overlap of the discrete state and the continuum. The exciton, inherently distinct from the continuum by its binding energy, is typically well separated. The necessary energetic overlap thus requires a significant spectral broadening of these states. Although the polydispersity of the sample may cause the broadening in the excitonic line's linewidth due to its inhomogeneity, we believe that this will not induce Fano interference. This phenomenon relies on coherence, and in solutions, inter-dot coherent coupling is anticipated to be minimal. Alternatively, we consider that the broadening is a result of thermal dephasing due to strong electron-phonon coupling. 
Interestingly, one can expect that the presence of Fano interactions further adds to the exciton dephasing and acts as a source of decoherence. Kuznetsova~\textit{et al.} discussed the effect of disorder in the homogeneous linewidth utilizing a Fano interference~\cite{Kuznetsova2010}.  Broido \textit{et al.}~\cite{Broido1998} also report a similar scenario in GaAs/Al$_{x}$Ga$_{1-x}$As quantum wells, where broadening of the excitonic features is induced by the Coulomb interactions of the overlapping continuum of states~\cite{Broido1990}. 

\section*{Supplementary Materials}

The supplementary materials includes the experimental methods for the perovskite nanocrystals preparation and its spectroscopic characterization (Two-dimensional photoluminescence excitation spectroscopy). Additionally, it shows the analysis to discard incoherent mixing in the experiments and supporting 2D-PLE spectra.

\begin{acknowledgments}
CSA acknowledges funding from the Government of Canada (Canada Excellence Research Chair CERC-2022-00055) and the Institut Courtois, Facult\'e des arts et des sciences, Universit\'e de Montr\'eal (Chaire de Recherche de l'Institut Courtois). 
ARSK acknowledges start-up funds from Wake Forest University, funding from the Center for Functional Materials, and funding from the Office of Research and Sponsored Programs at Wake Forest University. 
ERG and CSA acknowledge support from the National Science Foundation (DMR-2019444: Science and Technology Center for Integration of Modern Optoelectronic Materials on Demand).
\end{acknowledgments}


\section*{Author Declarations}
\subsection*{Conflicts of Interest}
The authors have no conflicts to disclose.

\subsection*{Author Contributions}
QA and LM synthesized and characterized the materials discussed in this manuscript. ARSK collected the 2D-PLE data and ERG carried out the primary analysis and modelling. ARSK and CSA led the intellectual conceptualization and co-supervised the project, and are corresponding co-authors. All co-authors contributed to the redaction of the manuscript. 

\section*{Data Availability}
The data that support the findings of this study are available from the corresponding co-authors upon reasonable request.


%


%



\end{document}